\documentclass[aps,prb,showpacs,floatfix,floats,superscriptaddress,twocolumn]{revtex4}
\usepackage{times}
\usepackage{bm}
\usepackage{subfig}
\usepackage{graphicx}
\newcommand{\be}{\begin{equation}}
\newcommand{\ee}{  \end{equation}}
\newcommand{\ba}{\begin{eqnarray}}
\newcommand{\ea}{  \end{eqnarray}}
\newcommand{\bi}{\begin{itemize}}
\newcommand{\ei}{  \end{itemize}}

\begin{document}

\title{Conductance quantization and transport gaps in disordered
graphene nanoribbons}

\author{E. R. Mucciolo}
\affiliation{Department of Physics, University of Central Florida, P.O. Box 
162385, Orlando, Florida 32816, USA}
\author{A. H. Castro Neto}
\affiliation{Department of Physics, Boston University, 590 Commonwealth 
Avenue, Boston, Massachusetts 02215, USA}
\author{C. H. Lewenkopf}
\affiliation{Departamento de F\'{\i}sica Te\'orica, Universidade do Estado 
do Rio de Janeiro, 20550-900 Rio de Janeiro, Brazil}
\affiliation{Laborat\'orio Nacional de Luz S\'{\i}ncrotron, Caixa Postal 6192, 13084-971 Campinas - S\~ao Paulo, Brazil}

\date{\today}

\begin{abstract}
We study numerically the effects of edge and bulk disorder on the
conductance of graphene nanoribbons. We compute the conductance
suppression due to Anderson localization induced by edge scattering
and find that even for weak edge roughness, conductance steps are
suppressed and transport gaps are induced. These gaps are
approximately inversely proportional to the nanoribbon width. On/off
conductance ratios grow exponentially with the nanoribbon length. Our
results impose severe limitations to the use of graphene in ballistic
nanowires.
\end{abstract}


\maketitle

\section{Introduction}
\label{sec:intro}

Measurements of electronic transport in graphene triggered an intense
effort to understand the physical properties of this
material.\cite{geim_review,RMPCastroNeto} Both fundamental and applied
aspects are currently being investigated by a large number of groups
around the world. As the quality of the samples improves and other
synthesis techniques are developed, the material changes from a regime
where bulk disorder is the dominant electron-scattering mechanism at
low temperatures to a ballistic one, where boundary conditions,
crystal alignment, and edge defects play a dominant role in setting
the transport properties. This regime is now experimentally accessible
in ultra narrow ribbons, which are promising for developing
high-frequency, low-noise, low-power field-effect transistors.


Motivated by recent experiments \cite{wang08,lin08,todd08} and
theoretical
studies,\cite{areshkin07,martin07,yan07,son07,rojas06,robinson07,li08,querlioz08,wakabayashi08}
in this paper we explore how edge roughness affects conductance in
long nanoribbons with realistic widths, from several nanometers to
tens of nanometers. Through numerical simulations we show that even
very weak edge disorder has a marked effect in the conductance of
these nanoribbons. Moderate amounts of edge roughness can
substantially suppress the linear conductance near the charge
neutrality point and induce a transport gap when the nanoribbon is
long and the number of propagating channels is small. This effect is a
manifestation of quasi-one-dimensional Anderson localization. We
compute transport gaps, localization lengths, and on/off ratios, and
explore the combined effect of edge and bulk disorders. We also
comment on the effect of inelastic scattering and dephasing. For
nanoribbons with very weak edge roughness, conductance steps appear at
noninteger values of the conductance quantum $e^2/h$, irrespective of
the lattice orientation. Our results indicate that producing quantum
point contacts with current graphene nanoribbons will be extremely
challenging and can only be achieved if either scattering at edges of
the constriction is substantially suppressed or the edges themselves
are defined with atomic precision. The paper is organized as
follows. In Sec. \ref{sec:method} we describe the numerical method and
present our results for the linear conductance of a graphene
nanoribbon in the presence of edge disorder. In Sec. \ref{sec:bulk} we
show the results of our computations when bulk disorder is added to
the nanoribbons and in Sec. \ref{sec:quantization} we discuss the
effects of weak edge roughness on the conductance quantization steps
of nearly ballistic nanoribbons. Conclusions and final comments are
left to Sec. \ref{sec:conclusions}.

\section{Transport in the presence of edge disorder}
\label{sec:method}

\begin{figure}[hb]
\centering
\label{fig:layout}
\includegraphics[width=0.8\columnwidth]{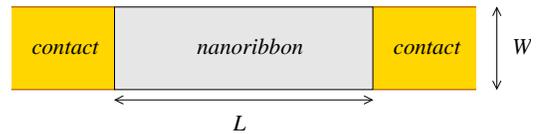}
\captionsetup{justification=RaggedRight,singlelinecheck=false}
\caption{(Color online) Schematic representation of the graphene
nanoribbon setup used in the simulations.}
\label{fig:ribbon}
\end{figure}

The simulations are based on the standard nearest-neighbor
tight-binding model of a single-layer graphene.\cite{RMPCastroNeto}
The linear conductance is evaluated through the recursive
Green's-function technique.\cite{Baranger91} An infinite nanoribbon is
broken into three parts (see Fig. \ref{fig:ribbon}): two (left and
right) semi-infinite regions of width $W$ modeling ideal contacts and
a finite central region of length $L$ where edge and bulk disorders
are introduced.

For a nanoribbon with perfect edges and no bulk
disorder,\cite{peres06} the conductance near the neutrality point
$E_F=0$ can be zero (for semiconductor armchair) or a multiple of the
conductance quantum $2e^2/h$ (for metallic armchair and zigzag),
depending on the availability of states. The first discontinuity in
the conductance appears when the Fermi energy $E_F$ matches the
minimum of first electron or hole subband; other steps are reached as
the minima of consecutive subbands are crossed.

In our simulations, edge defects are created by extracting lattice
sites (carbon atoms) from both edges of the nanoribbon following a
uniform probability distribution. It is assumed that atoms at the
edges are always attached to two other carbon atoms and passivated by
a neutral chemical ligand, such as hydrogen. The control parameters
are the number of etching sweeps $N_{\rm sweep}$, which is related to
the roughness amplitude, and the etching probability per site in the
$k$th sweep, $p_k$, which is related to the edge defect
density. Unless otherwise specified, an averaging over ten
realizations of each disorder case analyzed was carried out to
decrease sample-to-sample fluctuations and facilitate the
visualization of the results.

\begin{figure}[ht]
\centering
\includegraphics[width=\columnwidth]{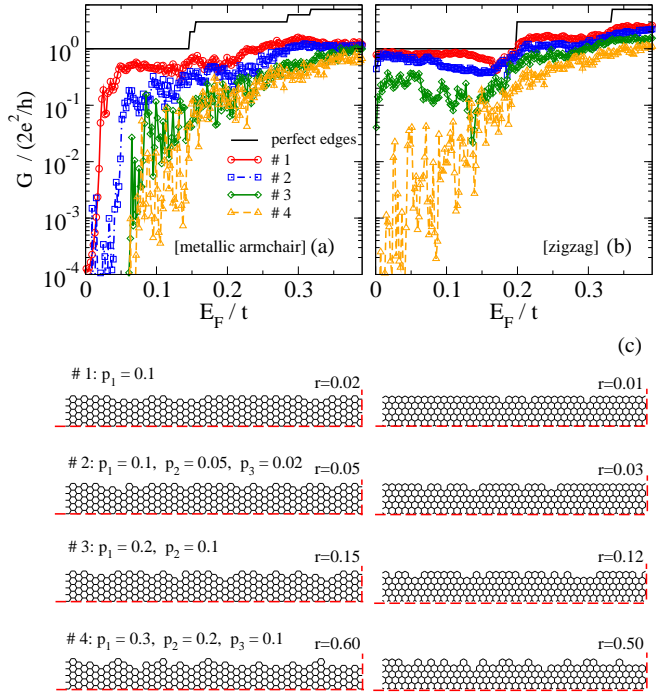}
\captionsetup{justification=RaggedRight,singlelinecheck=false}
\caption{(Color online) (a) and (b): Energy dependence of the average
linear conductance of graphene nanoribbons with varying edge
roughness. All nanoribbons have a the same length ($L = 45$ nm) and
similar widths ($W=4.4$ nm for armchair and $W=4.7$ nm for
zigzag). (c) Typical etching profiles used in (a) and (b) (only
segments of the nanoribbon atomic structure are shown). Left:
armchair and right: zigzag.}
\label{fig:cond_comp}
\end{figure}

In Figs. \ref{fig:cond_comp}a and \ref{fig:cond_comp}b we show the
linear conductance of nanoribbons with metallic armchair and zigzag
lattice orientations as a function of the Fermi energy for different
values of $N_{\rm sweep}$ and $\{p_k\}$ at zero temperature. Also
shown in Fig. \ref{fig:cond_comp}c are the typical etching profiles
for the four different cases. We have defined the roughness parameter
$r = (W - \overline{W})P/a_0$, where $\overline{W}$ is the average
ribbon width, $P=\sum_k p_k$, and $a_0$ is the lattice
constant.\cite{footnote} While a staircase of conductance steps is
seen in the absence of edge roughness, the conductance rapidly
degrades as the concentration and depth of the random etching
increase. The conductance is strongly suppressed near the neutrality
point even for relatively shallow etchings. Close to the neutrality
point, a deep gap develops.

The formation of a transport gap in ultranarrow graphene nanoribbons
is very much consistent with all experimental evidence available so
far.\cite{wang08,han,chen} Several mechanisms have been proposed to
account for this phenomenon, such as straightforward lateral
confinement to many-body effects. \cite{son06,antonioCB} Some of these
mechanisms require the existence of substantial edge disorder along
the ribbon (enough, for instance, to form bottle necks and quantum
dots.\cite{antonioCB})

The results presented in Figs. \ref{fig:cond_comp}b and
\ref{fig:cond_comp}c show that the suppression of conductance can also
occur at small values of $r$. Near the neutrality point, the number of
propagating channels in the nanoribbon is very small and the system
behaves as a quasi-one dimensional wire. Edge defects act as randomly
positioned short-range scatterers and induce strong backscattering,
which in turn leads to Anderson localization if the nanoribbon is
longer than the localization length. Note that even at room
temperatures, we expect dephasing lengths in graphene to be
exceedingly long.\cite{raman_exp} Therefore, in practice, localization
lengths can be shorter than both the nanoribbon length and the
dephasing length.

\begin{figure}[t]
\includegraphics[width=0.9\columnwidth]{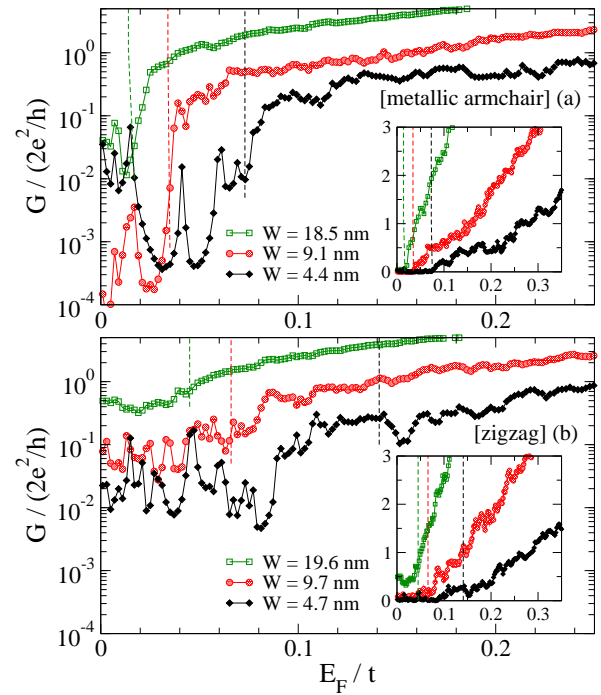}
\captionsetup{justification=RaggedRight,singlelinecheck=false}
\caption{(Color online) Average conductance of (a) metallic armchair
($W=4.3$ nm) and (b) zigzag ($W=4.6$ nm) nanoribbons as a function of
the Fermi energy for three different widths when moderate roughness is
present (zero temperature). The dashed lines indicate the conductance
gap estimated by the change in the slope of the curves (see insets). A
total of ten realizations of the edge roughness types \#3 for armchair
and \#4 for zigzag were used for each curve presented.}
\label{fig:cond_width_length}
\end{figure}

\begin{figure}[t]
\includegraphics[width=\columnwidth]{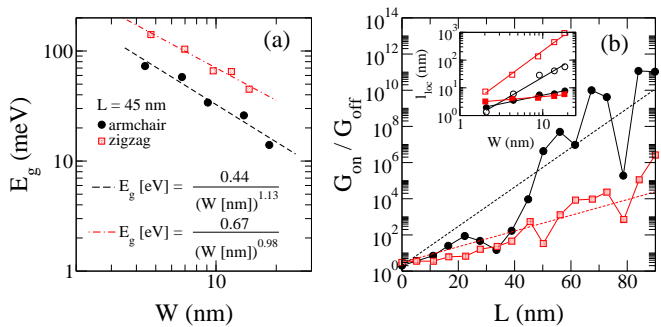}
\captionsetup{justification=RaggedRight,singlelinecheck=false}
\caption{(a): Transport gap dependence on the nanoribbon width. The
dashed and dashed-dotted lines are fittings to the data as explained
in the main text. (b) On/off linear conductance ratio as a function of
ribbon length. The solid lines are guides for the eyes and the dashed
lines indicate fittings to exponential curves. Each data point
corresponds to an average over 50 realizations of edge
roughness. Inset: localization length as a function of ribbon width
(circles are for armchair and squares are for zigzag). Solid (empty)
symbols represent $E_F = 0$ ($E_F = 0.2\,t$). Lines are fittings of
the functional form $\xi = A\, W^\alpha$ (see main text).}
\label{fig:gaps}
\end{figure}

Further evidence of this effect is provided in
Figs. \ref{fig:cond_width_length}a and \ref{fig:cond_width_length}b
where the conductance as a function of Fermi energy is shown for three
different nanoribbon widths. At a fixed energy, for nanoribbons with
increasing width, the localization becomes weaker and, consequently,
the conductance suppression decreases. In order to make contact with
experiments, we estimated transport gaps by determining the energy
value at which the curves show an inflection point [see insets of
Figs. \ref{fig:cond_width_length}a and
\ref{fig:cond_width_length}b]. This inflection point corresponds to
the crossing between two straight lines: an energy-independent
conductance at low energies and a linear function at high
energies. The results are shown in Fig. \ref{fig:gaps}a. For both
lattice orientations, we find that the transport gap $E_g$ scales
approximately with the inverse of the nanoribbon width $W$ and is only
weakly dependent on the length $L$ provided the latter is sufficiently
long (not shown). Notice that the value obtained for $A$, the scaling
prefactor, is in the same range of those found experimentally
($A\approx 0.2-0.6$ eV$\cdot$nm) (Refs. \onlinecite{han} and
\onlinecite{chen}) even for moderate roughness. The gap is less
pronounced for zigzag nanoribbons, since for this orientation most of
the current at low doping is carried through bulk states
\cite{areshkin07} which are less sensitive to edge
defects. Nevertheless, once the etching goes deeper than one lattice
spacing, a clear gap develops for zigzag orientations as well.

In Fig. \ref{fig:gaps}b we plotted the ratio between the
on and off linear conductances as a function of the nanoribbon
length. The off conductance $G_{\rm off}$ was obtained at $E_F=0$
while $G_{\rm on}$ was defined as the conductance at the Fermi energy
where a transition from the first to the second step occurs in the
clean nanoribbon. The curves indicate an approximate exponential
growth in the on/off ratio, consistent with the Anderson localization
picture. Since conductance is broadly distributed in the localized
regime, the on/off ratio develops very large fluctuations when the
nanoribbon is long. In the inset we show the localization length
$l_{\rm loc}$ extracted by fitting to the data an expression of the
form $G(L) = G(0)\, e^{-L/l_{\rm loc}}$. We note that the localization
length grows with increasing ribbon width and energy. Fittings to the
form $l_{\rm loc} = A\, W^\alpha$ yield $\alpha = 0.21-0.61$ at
$E_F=0$ and $\alpha = 1.8-2.2$ at $E_F=0.2\, t$, with the prefactor in
the range $A = 0.04-7.3$ for $l_{\rm loc}$ and $W$ in lattice
units. Thus, $l_{\rm loc}$ can be comparable to $W$.

\begin{figure}[t]
\includegraphics[width=\columnwidth]{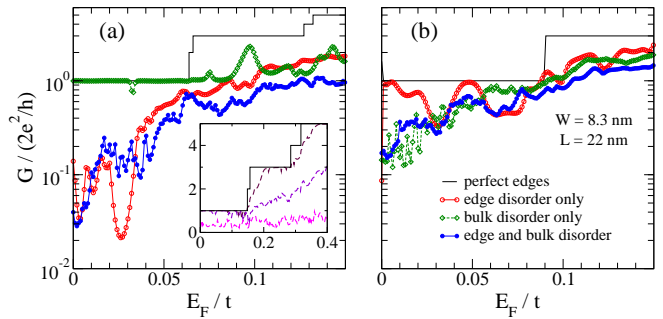}
\captionsetup{justification=RaggedRight,singlelinecheck=false}
\caption{(Color online) The effect of combined edge and bulk disorders
on the average conductance of short graphene nanoribbons. (a) Armchair
and (b) zigzag orientations (both with the same width $W = 8.3$ nm and
length $L = 22$ nm). The edge disorder corresponds to the case \#2 of
Fig. \ref{fig:cond_comp}c; the bulk disorder has parameters $n_{\rm
imp} = 0.04$, $K_0 = 0.5$, and $\xi = 2\, a_0$. Inset: average linear
conductance of armchair nanoribbons with bulk disorder and perfect
edges at zero temperature ($L = 22$ nm, $W = 4.4$ nm, and five
realizations per curve). The dashed lines correspond to correlation
lengths $\xi/a_0 =$ 10, 3, and 1 from top to bottom ($n_{\rm
imp}=0.02$ and $K_0=1$ in all three cases).}
\label{fig:cond_dis}
\end{figure}

\section{Bulk Disorder}
\label{sec:bulk}

In order to investigate the effects of bulk disorder, we added an
on-site correlated Gaussian disordered potential $U_{\rm imp}({\bf
r})$ to the nanoribbon.\cite{ShonAndo98,Rycerz06,lewenkopf08} The
latter is constructed by distributing along the nanoribbon $N_{\rm
imp}$ Gaussian scatterers of width $\xi$ with random amplitudes
$\{U_n\}$ drawn from a uniform distribution $[-\delta U,\delta
U]$. The intensity of the disorder is characterized by the
dimensionless parameter $K_0$, which is defined through the
correlation function
\begin{equation}
\label{eq:disorder}
\langle U_{\rm imp}({\bf r}_i) U_{\rm imp}({\bf r}_j) \rangle = K_0
\frac{\hbar v}{2\pi \xi^2} e^{-|{\bf r}_i - {\bf r}_j|^2/2\xi^2}.
\end{equation}
In the dilute limit, when $N_{\rm imp}$ is much smaller than the total
number of sites in the nanoribbon, one can show that \cite{Rycerz06}
$K_0 \approx 40.5\, n_{\rm imp} (\delta U/t)^2 (\xi/a_0)^4$, where
$n_{\rm imp}$ is the scatterer density per lattice site. For large
graphene sheets at high doping, far from the neutrality point, it is
possible to relate this parameter to the transport mean-free path
using the Born approximation:\cite{ShonAndo98} $\ell_{\rm tr} =
2\lambda_F/(\pi K_0)$, where $\lambda_F$ is the Fermi wavelength in
the graphene sheet, with $\lambda_F \ll \ell_{\rm tr}$.

The difference between armchair and zigzag states near the neutrality
point also explains the effect of bulk disorder in nanoribbons with
rough edges (Fig. \ref{fig:cond_dis}). For the metallic armchair
orientation the low-lying states are concentrated at the edges and are
quite sensitive to edge roughness.\cite{areshkin07} In this case,
moderate bulk disorder has a small effect on the transport gap. For
zigzag orientations, the situation is the opposite, as bulk disorder
disrupts the current-carrying states and suppresses conductance in
this case. However, for both orientations, bulk disorder only leads to
strong localization when $\xi \alt a_0$ (short-range disorder).

\begin{figure}[t]
\includegraphics[width=\columnwidth]{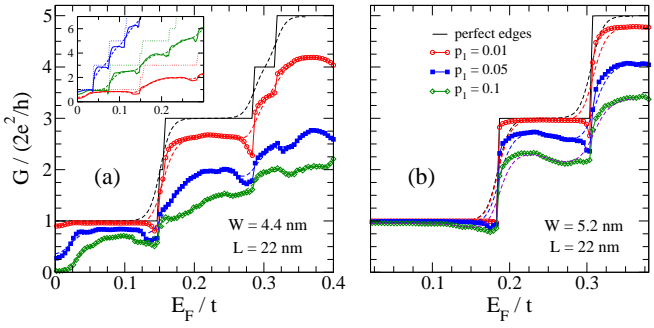}
\captionsetup{justification=RaggedRight,singlelinecheck=false}
\caption{(Color online) Suppression of conductance quantization steps
due to weak edge disorder. (a) Armchair and (b) zigzag orientations. A
total of 100 realizations were used in each case. Solid (dashed) lines
indicate zero (room) temperature in the contacts. Inset: average
conductance versus energy for a metallic armchair nanoribbon with
$L=22$ nm, no bulk disorder, $p_1 = 0.05$ (ten realizations), and
increasing widths: $W = 4.4$, 9.1, and 18.5 nm; solid (dashed) lines
represent zero (room) temperature, while dotted lines describe the
perfect edge case at zero temperature.}
\label{fig:cond_sup}
\end{figure}

\section{Conductance Quantization}
\label{sec:quantization}

While it is clear that moderate edge disorder leads to substantial
suppression of the conductance of nanoribbons, what happens when the
etching is nearly perfect and only very few and shallow defects are
present? How does weak edge roughness affect conductance quantization
in comparison to bulk disorder? Answers to these questions are
provided in Fig. \ref{fig:cond_sup} and in the inset of
Fig. \ref{fig:cond_dis}a.  The main effect of very weak edge disorder
is to lower the conductance steps {\it without} changing their
positions in energy. Some fluctuations occur near the transition
regions between steps because of the sensitivity of the evanescent
modes to variations in nanoribbon width, but these fluctuations are
washed out at finite temperatures. However, thermal fluctuations, even
at room temperature, do not change the wider steps, as can be seen in
the insets of Fig. \ref{fig:cond_sup}, as long as the nanoribbon is
sufficiently narrow.

Bulk disorder, on the other hand, has a quite distinct effect on the
conductance quantization. In the inset of Fig. \ref{fig:cond_dis}(a)
we show how bulk disorder affects the conductance steps of a metallic
armchair nanoribbon. As the disorder range widens, the steps are
smeared without shifting the conductance value. This result can be
understood in the following way. When only weak edge disorder is
present, the width of the nanoribbon is hardly unaltered and
propagating channels in the nanoribbon open up at the same energies as
in the case of perfect edges. Yet, backscattering due to randomly
positioned edge defects, albeit weak, reduces the overall conductance
and shifts the steps downward. Long-range bulk disorder, on the other
hand, creates potential inhomogeneities which lead to the appearance
of electron and hole puddles when the Fermi energy lies close to the
Dirac point.\cite{lewenkopf08} Transmission through these puddles
creates mode mixing, which in turn smears the conductance steps.

A simple model can be used to describe the suppression of the
conductance steps. For the first step, let us assume that carriers
propagate in one dimension through a sequence of randomly positioned
but identical barriers. When the barrier reflectance is very small,
$R\ll 1$, it is straightforward to show that, in the short-wave limit
$(\lambda \ll L)$, the conductance in the first step goes as $G_1
\approx (2e^2/h)\, [1-NR]$, where $N$ is the number of barriers, which
can be directly related to the defect probability or density as
follows: $N = (L/a_0)p_1$. For higher steps, more than one propagating
mode is present and the system becomes quasi-one dimensional. In the
absence of mode mixing, the conductance of the $n$th step behaves as
$G_n \approx n\, (2e^2/h)\, [1-NR]$. However, mode mixing is
unavoidable for $n>1$ and one expects strong deviations from this
simple scaling behavior

\begin{figure}[h]
\includegraphics[width=0.95\columnwidth]{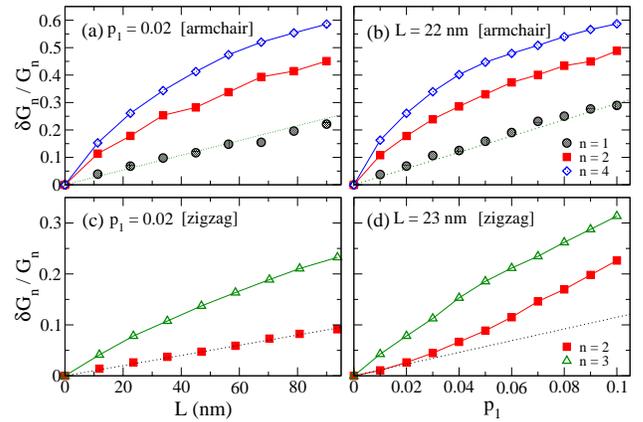}
\captionsetup{justification=RaggedRight,singlelinecheck=false}
\caption{(Color online) Relative suppression of the $n$th conductance
step as a function of (a) and (c) nanoribbon length and (b) and (d)
defect concentration for metallic armchair ($W=4.4$ nm) and zigzag
($W=5.2$ nm) orientations. Each data set corresponds to an average
over 100 realizations. The index $n$ indicates the order of the
conductance step and the solid lines are guides for the eyes. The
dotted lines are explained in the main text.}
\label{fig:dG_p_N}
\end{figure}

The suppression of the conductance step as a function of nanoribbon
length and edge defect concentration is presented in
Fig. \ref{fig:dG_p_N}. For the first step of the armchair orientation
the simple one-dimensional scattering model works quite well. Both
dashed lines in Figs. \ref{fig:dG_p_N}a and \ref{fig:dG_p_N}b
correspond to $R=0.035$. The linear scaling ceases to apply for higher
steps, with a sublinear dependence indicating substantial mode
mixing. For zigzag orientations, the suppression of the first
conductance step was too small to be shown on the same plot. Figures
\ref{fig:dG_p_N}c and \ref{fig:dG_p_N}d show that the simple scaling
behavior no longer applies already when $n=2$ ($R=0.013$ was used in
this case).

\section{Conclusions}
\label{sec:conclusions}

Our results indicate that creating graphene quantum point contacts
will depend fundamentally on atomic scale
engineering.\cite{tapaszto08} In searching for ways to improve the
nanoribbon conductance in the presence of edge roughness, we found
that, due to the Klein tunneling effect, side gates are not effective
in reducing edge scattering (i.e., electrostatic potentials do not
confine Dirac fermions). At moderate roughness, Anderson localization
develops and a transport gaps appear. There is a quantitative
agreement between the gaps that we find numerically when extrapolating
our results to wider nanoribbons and the available experimental data,
although we expect that other mechanisms, such as spin gaps
\cite{son06} and charging effects \cite{antonioCB} (not taken into
account in our calculations), will likely compete with
localization. Electronic dephasing can be introduced in the
calculation following a standard procedure.\cite{damato90} Since, in
practice, dephasing lengths exceed the nanoribbon width, we expect
that the main effect of dephasing will be the appearance of an
additional, weakly energy dependent, suppression of the conductance,
as the nanoribbon will break into a series of independent quantum
resistors. At this point, we would also like to note that we recently
became aware of similar work by Evaldsson {\it et
al}.\cite{evaldsson08}

\acknowledgments

We thank I. Adagideli, H. Dai, D. Goldhaber-Gordon, M. Ishigami,
Y.-M. Lin, I. Martin, and K. Todd for helpful discussions. E.R.M. and
A.H.C.N. thank the Aspen Center for Physics, where this work was
completed. C.H.L. thanks FAPERJ and CNPq (Brazil) for partial
financial support.



\begin{thebibliography}{99}
\bibitem{footnote} This is a sensible definition only when $P\ll 1$.
\end{thebibliography}
\end{document}